\documentclass{jfm}

\usepackage{graphicx}
\usepackage{natbib}

\ifCUPmtlplainloaded \else
  \checkfont{eurm10}
  \iffontfound
    \IfFileExists{upmath.sty}
      {\typeout{^^JFound AMS Euler Roman fonts on the system,
                   using the 'upmath' package.^^J}%
       \usepackage{upmath}}
      {\typeout{^^JFound AMS Euler Roman fonts on the system, but you
                   dont seem to have the}%
       \typeout{'upmath' package installed. JFM.cls can take advantage
                 of these fonts,^^Jif you use 'upmath' package.^^J}%
      }
  \else
  \fi
\fi

\ifCUPmtlplainloaded \else
  \checkfont{msam10}
  \iffontfound
    \IfFileExists{amssymb.sty}
      {\typeout{^^JFound AMS Symbol fonts on the system, using the
                'amssymb' package.^^J}%
       \usepackage{amssymb}%

      }{}
  \fi
\fi

\ifCUPmtlplainloaded \else
  \IfFileExists{amsbsy.sty}
    {\typeout{^^JFound the 'amsbsy' package on the system, using it.^^J}%
     \usepackage{amsbsy}}
    {\providecommand\boldsymbol[1]{\mbox{\boldmath $##1$}}}
\fi

\newsavebox{\astrutbox}
\sbox{\astrutbox}{\rule[-5pt]{0pt}{20pt}}

\usepackage{color}

\title[Wake pattern and wave resistance for  anisotropic moving objects]{Wake pattern and wave resistance for  anisotropic moving objects}

\author[M. Benzaquen, A. Darmon and E. Rapha\"el]%
{Michael Benzaquen$^1$, 
Alexandre Darmon$^{2}$ and Elie Rapha\"el$^1$\thanks{Email address for correspondence: elie.raphael@espci.fr} }

\affiliation{$^1$PCT, UMR CNRS 7083 Gulliver, ESPCI ParisTech, 10 rue Vauquelin, 75005 Paris, France\\[\affilskip]
$^2$EC2M, UMR CNRS 7083 Gulliver, ESPCI ParisTech, 10 rue Vauquelin, 75005 Paris, France}

\pubyear{2010}
\volume{650}
\pagerange{119--126}
\begin{document}

\maketitle

\begin{abstract}
We present a theoretical study of gravity waves generated by an anisotropic moving disturbance. We model the moving object by an elliptical pressure field of given aspect ratio $\mathcal W$. We study the wake pattern  as a function of $\mathcal W$ and the longitudinal hull Froude number $Fr = V/\sqrt{gL}$, where $V$ is the velocity, $g$ the acceleration of gravity and $L$ the size of the disturbance in the direction of motion.  For large hull Froude numbers, we analytically show that the rescaled surface profiles for which $\sqrt{\mathcal W}/Fr$ is kept constant coincide. In particular, the angle outside which the surface is essentially flat remains constant and equal to the Kelvin angle, and the angle corresponding to the maximum amplitude of the waves scales as $\sqrt{\mathcal W}/Fr$, thus showing that previous work on the wake's angle for isotropic objects can be extended to anisotropic objects of given aspect ratio.
We then focus on the wave resistance and discuss its properties in the case of an elliptical Gaussian pressure field. We derive an analytical expression for the wave resistance in the limit of very elongated objects, and show that the position of the speed corresponding to the maximum wave resistance  scales as $\sqrt{gL}/\sqrt{\mathcal W}$.
\end{abstract}

%

\section{Introduction}
The influence of Lord Kelvin's theory \citep{Kelvin1887} on the study of water waves and particularly on the wake pattern that falls behind an object moving at constant speed is no longer to be proved. His work revealed a certain universality of the wake pattern, that everyone can notice when looking at the waves produced by objects as different as swans or sailing boats. One of his most remarkable results concerns the well-known constant wake angle $\varphi_\mathrm{K} \simeq 19.47^{\circ}$, delimiting a region outside which the water remains essentially unperturbed. Since Lord Kelvin's first results, a lot of efforts have been done to extend his work \citep{Lighthill1978,Barnell1986,Lamb1993,Johnson1997}, for instance to nonlinear waves \citep{Dias1999}, or waves in the presence of vorticity \citep{Ellingsen2014,Ellingsen2014-2}.
Recently the constancy of the wake angle has been contested \citep{rabaud2013ship}. Indeed, in their analysis of airborne images, \textit{Rabaud and Moisy} showed that, for large hull Froude numbers $Fr=V/\sqrt{gL}$, where $V$ is the object velocity, $g$  the  acceleration of gravity and $L$  the typical size of the object in the direction of motion, the angle of the wake decreases and scales as $1/Fr$. In response to these intriguing  observations, we recently \citep{Darmon2014} presented an analytical study of the wake pattern as a function of the Froude number and showed that the delimiting angle of the wake actually remains constant for all $Fr$, therefore comforting Lord Kelvin's theory.  We also provided an explanation to the airborne observations of \textit{Rabaud and Moisy} by analytically proving that, for an axisymmetric object of typical size $L$,  the angle corresponding to the maximum amplitude of the waves decreases  as $Fr$ is increased, scaling as $1/Fr$ for large Froude numbers.  
\smallskip

The main issue that arose from  our previous study \citep{Darmon2014} is that an axisymmetric object does not correctly reflect the real geometry of boats \citep{Dias2014}. Indeed, boats have elongated shapes with aspect ratios typically ranging between 0.1 and 0.5, with the exception of overcrafts which actually display a cylindrical symmetry. For  elongated objects, the emphasis has been placed on the planing regime in the limit of small aspect ratios \citep{Casling1978,Taravella2011}. The important feature that needs to be considered when it comes to hull design, is the so-called wave resistance, or wave drag, a force resulting from the generation of surface waves. Indeed, a well profiled hull notably reduces the wave resistance thus improving the velocity performances of the ship. Along with wake patterns, wave resistance has been widely studied both at practical \citep{Peri2001,Darrigol2005} and theoretical levels \citep{Havelock1908,Wehausen1973,Lighthill1978}. 
\smallskip

We here present a theoretical study on anisotropic moving objects. In the first part, we focus on the wake pattern. After recalling the main ingredients of the physical model, we calculate the surface displacement induced by an elliptical moving pressure distribution and discuss its main features. In particular, we investigate the universality of the wake pattern as a function of the aspect ratio of the moving disturbance and the longitudinal hull Froude number. In the second part, we derive the expression of the wave resistance for an anisotropic moving disturbance and focus on the case of an elliptical  Gaussian pressure field. In the limit of very elongated objects, we obtain interesting analytical results on the scaling of the velocity corresponding to the maximum of the wave resistance.



 
\section{Wake pattern}

\begin{figure}
  \centering \includegraphics[scale=0.388]{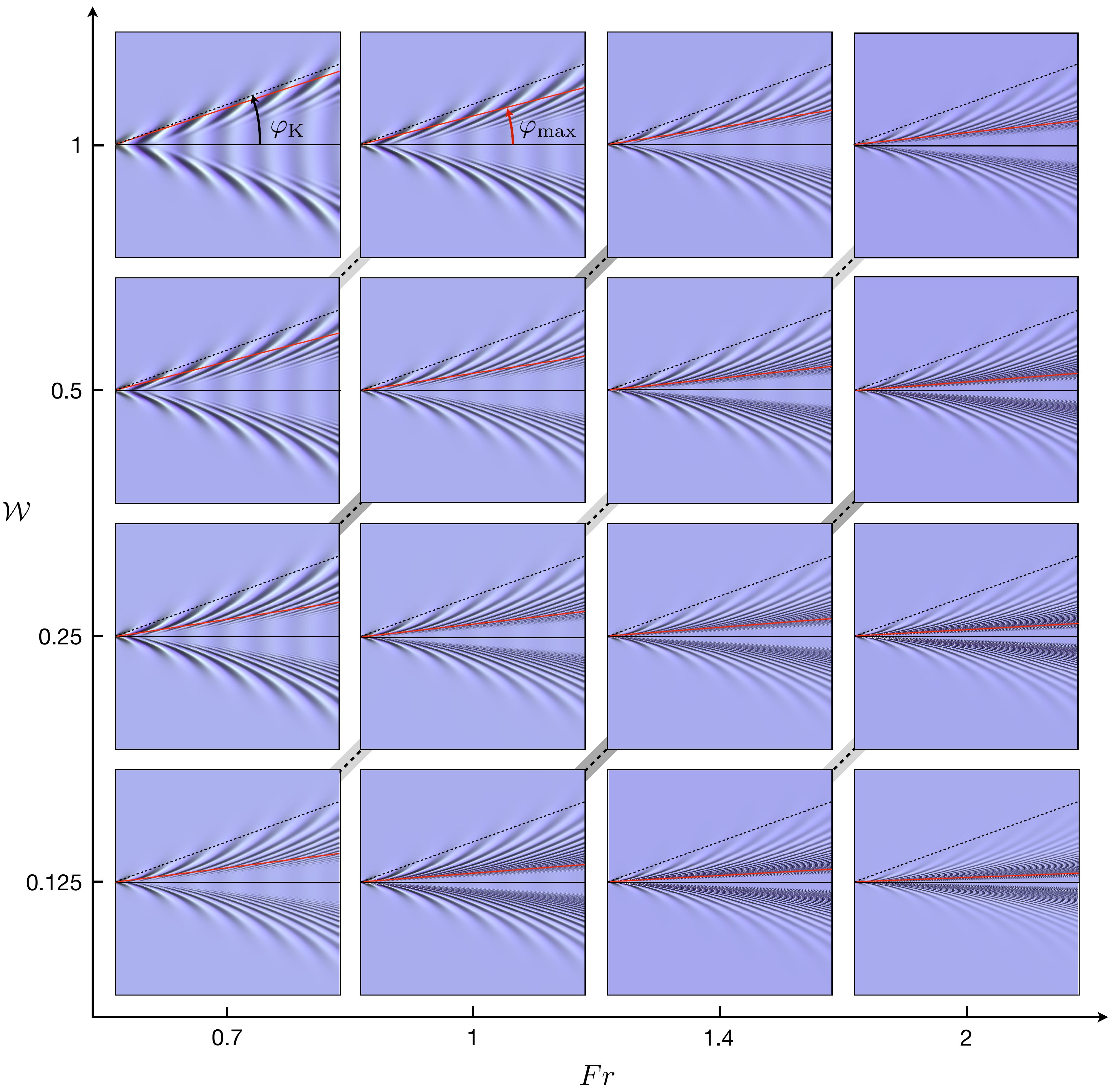}
 \caption{Colour online. Relief plots of the rescaled surface displacement produced by a moving elliptical Gaussian pressure field (see Eq.~(\ref{Gauss})) computed using Eq.~(\ref{zeta}). Here, rescaled means multiplied by $\sqrt{\mathcal W}$. Each relief plot corresponds to given values of the longitudinal hull Froude number $Fr$ and aspect ratio  $\mathcal W$ and is plotted as a function of $\tilde X=X/\Lambda$  and $\tilde Y=Y/\Lambda$ where $\Lambda=2\pi Fr^2$ is the dimensionless wavelength. The parameters $Fr$ and $\mathcal W$ are varied respectively along the horizontal axis and the vertical axis on a logarithmic scale. In all graphs the angle $\varphi_{\mathrm{max}}$ indicating the maximum amplitude of the waves is signified with a solid red line and the Kelvin angle $\varphi_{\mathrm K}=19,47^\circ$ with a black dashed line.
On each  diagonal, signified by a grey stripe, the value of the ratio $\mathcal U = \sqrt{\mathcal W} \,  Fr^{-1}$ is kept constant.
The first row ($\mathcal W=1$) corresponds to the isotropic case.}
\label{profils}
\end{figure}

For gravity waves, the surface displacement generated by a pressure field $p(x,y)$ moving in the $-x$ direction with constant speed $V$ can be written in the frame of reference of the moving perturbation as \citep{Havelock1908,Havelock1919,Darmon2014}:
\begin{eqnarray}
\zeta(x,y) &=& -\lim_{\varepsilon\rightarrow 0}   \, \int \!\!\!\! \int \frac{\mathrm{d} k_x \, \mathrm{d} k_y}{4\pi^2\rho}  \, \frac{k\,\hat{p}(k_x,k_y) \, e^{-i(k_x x+k_y y)}}{\omega(k)^2-V^2k_x^2+   2 i \varepsilon Vk_x } \, , 
\label{profil12}
\end{eqnarray}
where  $k=\left(k_x^2+k_y^2\right)^{1/2}$, $\hat{p}(k_x,k_y)$ is the Fourier Transform of $p(x,y)$, $\rho$ is the water density and $\omega(k)=(gk)^{1/2}$  is the dispersion relation for pure gravity waves. Let us nondimensionalise the problem through $x=LX  \, ; \,\, y=LY\, ; \, \,k_x= K_X/L \,;\, \,k_y= K_Y/L \,;\, \,k= K/L \,; \, \,
{p}={\rho g L}\,{{P}} \,; \, \, \displaystyle\hat{p}={\rho g L^3}{\hat{P}} \,;  \,\, {\varepsilon}={(g/L)^{1/2}}{\tilde\varepsilon}$ \ where $L$ is the typical size of the pressure field $p(x,y)$ in the direction of motion. This yields $\zeta(x,y)= ({L}/{4\pi^2})  \,Z(X,Y)$
where:
 \begin{eqnarray}
 Z(X,Y)&=&    -\lim_{\tilde\varepsilon\rightarrow 0}   \, \int \!\!\!\! \int {\mathrm{d} K_X \, \mathrm{d} K_Y}  \, \frac{K\hat{P}(K_X,K_Y) \, e^{-i(K_X X+K_Y Y)}}{K-Fr^2K_X^2+   2 i \tilde\varepsilon \,Fr\,K_X }  
     \ ,
\label{}
\end{eqnarray}
and where $Fr=V/\sqrt{gL}$ is the longitudinal hull Froude number.
In order to account for the anisotropic geometry of a real ship's hull in a simple manner, we consider an elliptical pressure field of the form: 
\begin{eqnarray}
P(X,Y)&=&F\left(X^2+\frac{Y^2}{\mathcal W^{2}}\right)
\label{ellip}
\end{eqnarray}
where $\mathcal W=l/L$ is the aspect ratio of the ellipse of major diameter $L$ and minor diameter $l$. Consistently with this choice, we let the elliptical change of variables: 
\begin{eqnarray}
X=\breve R\cos\breve \varphi &\,\,\,\,\,\,\,\,\,\,\,\,& K_X=\breve K\ \cos\breve\theta  \\ \label{elliptical}
Y=\mathcal W\breve R\sin\breve \varphi &&K_Y=\mathcal W^{-1}\breve K \sin\breve\theta\ . \nonumber
\label{}
\end{eqnarray}
This leads to  $Z(X,Y)=\breve Z(\breve R,\breve \varphi)$ where:
\begin{equation}
\breve Z(\breve R,\breve \varphi)=  -\lim_{\tilde\varepsilon\rightarrow 0} \,   \int \!\!\!\! \int    \breve K\mathrm  d \breve K \mathrm  d \breve \theta \,   \frac{    \sqrt{\mathcal W^{2} \cos^2\breve \theta+   \sin^2 \breve \theta}  \, \breve{\hat P} (\breve K)\,e^{-i\breve K\breve R\cos(\breve \theta-\breve \varphi)}}{    \sqrt{\mathcal W^{2} \cos^2\breve \theta+  \sin^2 \breve \theta} -     \mathcal V^2\breve K\cos^2\breve \theta +2i\tilde \varepsilon\,\mathcal V\sqrt{\mathcal W} \cos{\breve \theta}}\ ,
\label{}
\end{equation}
and where $\mathcal V=Fr\sqrt{\mathcal  W}$ and $ \breve{\hat P} (\breve K)=\mathcal W^{-1}\hat P (K_X,K_Y)$. The interest of the elliptical change of variables in Eq.~(\ref{elliptical}) is that the Fourier transform of the pressure field is now a function of the single variable $\breve K$. This, as we shall see later, implies that our previous derivation with an axisymmetric pressure field \citep{Darmon2014} is applicable to the case of an elliptical pressure field. Using the Sokhotski-Plemelj formula (see \textit{e.g.} \citep{Appel2007}) to perform the integral over $\breve K$ yields $\breve Z(\breve R,\breve \varphi)=\breve Z_0(\breve R,\breve \varphi)+\breve G(\breve R,\breve \varphi)$  where $\breve G(\breve R,\breve \varphi)$ is a rapidly decreasing function with the distance to the perturbation and where:
\begin{eqnarray}
\breve Z_0(\breve R,\breve \varphi)&=&{i\pi}\int_{-\pi/2}^{\pi/2}  {\mathrm{d} \breve \theta} \,\, \frac{(\mathcal W^{2} \cos^2\breve \theta+   \sin^2 \breve \theta)\,{\breve{\hat{P}}(\breve K_0) \, e^{-i\breve K_0 \breve R \cos (\breve \theta -\breve \varphi)}}  }{\mathcal V^4\cos^4 \breve \theta}    \, ,\label{zeta}
\end{eqnarray}
where $\breve K_0=\sqrt{ \mathcal W^2 \cos^2\breve \theta+  \sin^2 \breve \theta}/({\mathcal V^2\cos^2\breve \theta})$. 
 The integral in Eq. (\ref{zeta}) is of the form 
   $\int \mathrm{d} \breve \theta   H(\breve \theta)\,e^{i\, \phi(\breve \theta)}$ and may be approximated through the method of the steepest descent.
 For $\breve R/\mathcal V^2>1$, the integrand oscillates rapidly and there are two stationary points $\breve \theta_1$ and $\breve \theta_2$ given by $\phi'(\breve \theta)=0$. 
One can then write $\breve Z_0(\breve R,\breve \varphi)\simeq i\pi(\breve Z_{01}(\breve R,\breve \varphi)+\breve Z_{02}(\breve R,\breve \varphi))$,
where $\breve Z_{01}$ corresponds to the transverse waves and $\breve Z_{02}$ corresponds to the diverging waves. For large Froude numbers, the transverse waves vanish compared to the diverging waves \citep{Darmon2014}, so that within  this limit we shall only consider the latter:
\begin{eqnarray}
\breve Z_{02}(\breve R,\breve \varphi)&=&\breve  Q(\breve R,\breve \varphi) \, e^{i\left(\phi(\breve R,\breve\theta_{2}(\breve \varphi),\breve \varphi)+\frac \pi4  \right)}  \, , \label{Z2}  
\end{eqnarray}
where $\breve Q(\breve R,\breve \varphi)$ is the envelope of the wave signal:
\begin{eqnarray}
\breve Q(\breve R,\breve \varphi)&=&   
 \sqrt{  \frac{2\pi}{\left|\partial_{\breve \theta}^2\phi(\breve R,\breve\theta_{2}(\breve \varphi),\breve \varphi)\right|}   } \, H(\breve \theta_{2}) \ .
\end{eqnarray}
For small elliptical angles $\breve \varphi$, the stationary point $\breve \theta_2$ reads 
$\breve \theta_2(\breve \varphi)=  {\pi}/{2} +2\breve \varphi$, and
the enveloppe $\breve Q(\breve R,\breve \varphi)$ of the waves eventually becomes in the small $\breve \varphi$ approximation:
\begin{eqnarray}
\breve Q(\breve R,\breve \varphi)&\simeq&\frac{1}{4}\sqrt\frac{\pi}{{\breve R}}    \frac{1}{\mathcal V^3\,\breve \varphi^{5/2}}\,\breve{\hat P}\left( \frac{1}{4\,\mathcal{V}^2 \breve \varphi^2}  \right) \ . \label{breveQ}
\end{eqnarray}
It is worth mentioning here that we recover a very similar expression to that of the isotropic case \citep{Darmon2014}, where the hull Froude number $Fr$ has been replaced by $\mathcal V$, and $\varphi$ by $\breve \varphi$. Let us now change the variables  back to real polar coordinates $(R,\varphi)$: 
\begin{subeqnarray}
R&=&\breve R\left(\cos^2\breve \varphi+\mathcal W^2 \sin^2\breve \varphi\right)^{1/2} \\
 \varphi&=& \arctan (\mathcal W \tan \breve \varphi) \ .\label{CDVAlex}
\end{subeqnarray}
For small angles, Eqs.~(\ref{CDVAlex}) become:
\begin{subeqnarray}
R&\simeq&\breve R \\ \label{realcoor}
 \varphi&\simeq& \mathcal W  \breve \varphi \ .
\end{subeqnarray}
Combining Eq.~(\ref{breveQ}) and Eq.~(\ref{realcoor}) and defining the rescaled enveloppe in real polar coordinates $Q(R,\varphi)= \sqrt{\mathcal W} \ \breve Q(\breve R,\breve \varphi)$ one gets:
\begin{eqnarray}
Q(R,\varphi) &\simeq&\frac{1}{4} \sqrt{\frac{\pi}{R}}  \,   \frac{\mathcal U^3}{ \varphi^{5/2}}\,\breve{\hat P}\left( \frac{\mathcal U^2}{4  \varphi^2}  \right) \ ,
\end{eqnarray}
where $\mathcal U = \mathcal W \,\mathcal V^{-1} = \sqrt{\mathcal W} \,  Fr^{-1}$. This shows that the rescaled enveloppe $Q(R,\varphi)$ depends on the aspect ratio $\mathcal W$ and the rescaled Froude number $\mathcal V$ only through the single variable $\sqrt{\mathcal W} \,  Fr^{-1}$. In other terms, for large Froude numbers and small angles, profiles resulting from moving disturbances with the same $\sqrt{\mathcal W} \,  Fr^{-1}$ coincide when multiplied by $ \sqrt{\mathcal W}$ (see Fig.~\ref{profils}).  In particular the angle $\varphi_{\mathrm {max}}$ corresponding to the maximum amplitude of the waves reads:
\begin{eqnarray}
\varphi_{\mathrm {max}}&\sim&   \frac{\sqrt{\mathcal W}}{Fr}  \ .
\end{eqnarray}
This explains why objects with very low aspect ratio, such as rowing boats, usually  display a small maximum amplitude angle even though their longitudinal hull Froude number is not very high as they are propelled by man power.
\medskip

To illustrate these analytical results, we then perform a numerical evaluation of the integral in Eq.~(\ref{zeta}) with an elliptical Gaussian pressure field whose dimensionless Fourier transform reads:
\begin{eqnarray}
\hat P(K,\theta) = \frac{f_0}{\rho gL^3}\exp\left[-\frac{K^2}{4\pi^2} (\cos^2\theta + \mathcal W^2 \sin^2\theta) \right] \ , \label{Gauss}
\end{eqnarray}
where $f_0$ is the total integrated pressure force. The resulting rescaled surface profiles are displayed in Fig.~\ref{profils}. Here, rescaled means multiplied by $\sqrt{\mathcal W}$. Each relief plot corresponds to given values of $Fr$ and $\mathcal W$ and is plotted as a function of $\tilde X=X/\Lambda$  and $\tilde Y=Y/\Lambda$ where $\Lambda=2\pi Fr^2$ is the dimensionless wavelength. The parameters $Fr$ and $\mathcal W$ are varied respectively along the horizontal axis and the vertical axis on a logarithmic scale. In all graphs the angle $\varphi_{\mathrm{max}}$ indicating the maximum amplitude of the waves is signified with a solid red line and the Kelvin angle $\varphi_{\mathrm K}$ with a black dashed line.
On each  diagonal, signified by a grey stripe, the value of the ratio $\mathcal U = \sqrt{\mathcal W} \,  Fr^{-1}$ is kept constant. The first row ($\mathcal W=1$) corresponds to the isotropic case.
Firstly, one can see that for any set of parameters ($Fr$, $\mathcal W$) the angle delimiting the region outside which the surface remains essentially flat is constant and equal to the Kelvin angle $\varphi_{\mathrm K}$.
Secondly, on a given   diagonal ($ \sqrt{\mathcal W} \,  Fr^{-1}=cst$), all the profiles seem identical to the naked eye, therefore comforting the analytical results found in the previous section. Finally, looking closely, one can note that the agreement between different profiles on the same diagonal becomes all the more true as we go towards higher values of $Fr$, as predicted by the analytical study.

\pagebreak

\section{Wave Resistance}

\begin{figure}
  \centering \includegraphics[scale=0.35]{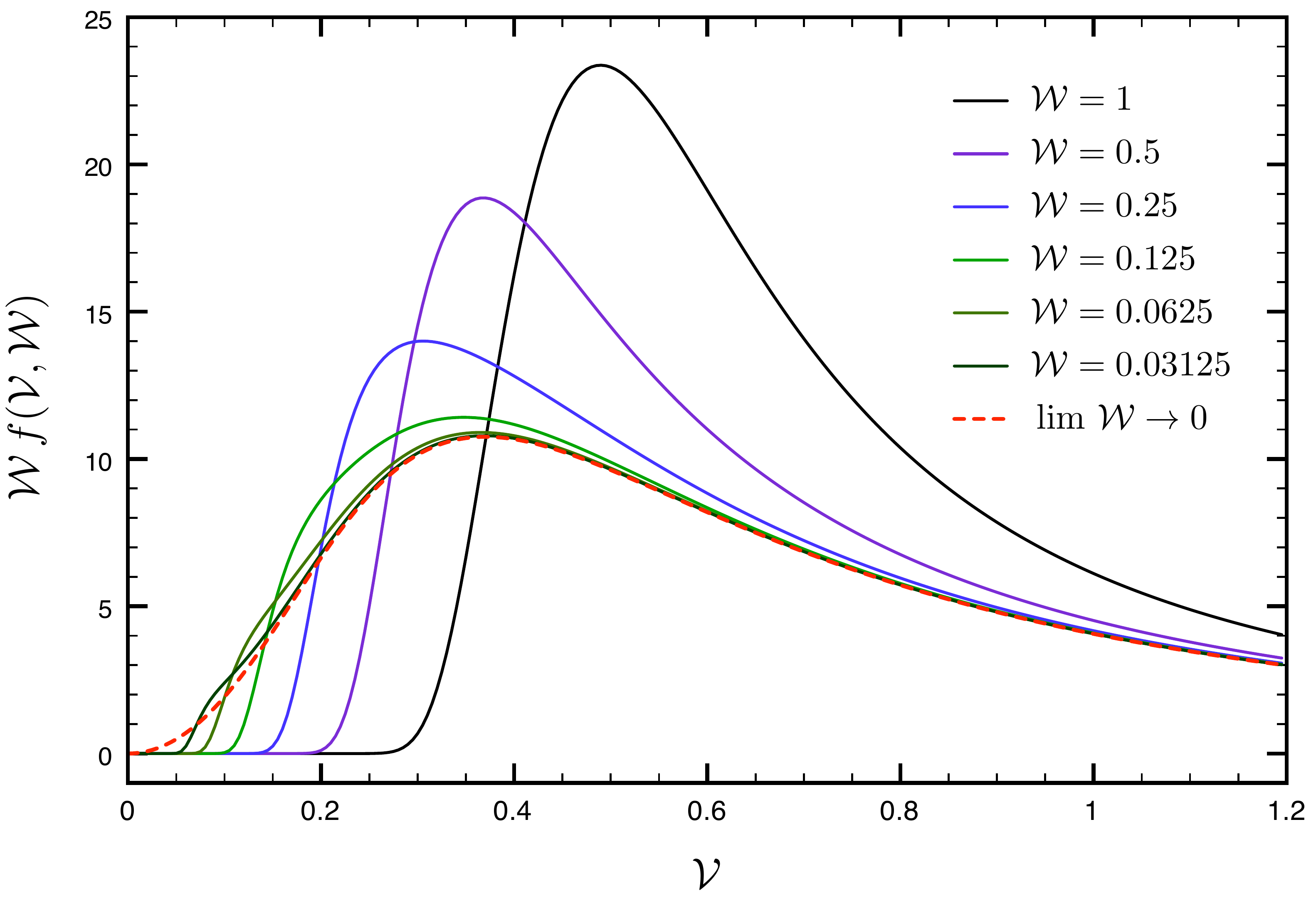}
 \caption{Colour online. Plot of $\mathcal W\,f(\mathcal V, \mathcal W)$ as given by Eq. (\ref{erfc})  as a function of the dimensionless parameter $\mathcal V=Fr\sqrt{\mathcal W}$, where $Fr=V\sqrt{gL}$ is the longitudinal hull Froude number,  for different values of the aspect ratio $\mathcal W$. The red dashed curve corresponds to the limit regime $\mathcal W \rightarrow 0$ as given analytically  by Eq.~(\ref{lim}).}
\label{resvag}
\end{figure}

The waves generated by the moving disturbance continually remove energy to infinity.  This translates into a  drag force $R$ experienced by the disturbance, often referred to as the wave drag or the wave resistance.  
We here investigate the wave resistance experienced by the disturbance, using the method proposed by Havelock \citep{Havelock1908,Havelock1919}, according to whom the wave resistance is the total resolved pressure in the direction of motion:
\begin{eqnarray}
R&=&- \int \!\!\!\! \int  \mathrm  dx\,\mathrm  d y\,p(x,y)\,\partial_x\zeta (x,y)\ . \label{Havelock}
\end{eqnarray}
Inserting Eq.~(\ref{profil12}) into Eq.~(\ref{Havelock}) yields \citep{Raphael1996}:
\begin{eqnarray}
R&=&\lim_{\varepsilon\rightarrow 0}\int \!\!\!\! \int\frac{\mathrm{d} k_x\, \mathrm{d}k_y}{4\pi^2\rho}\,\frac{ik_x k\left|\,\hat p(k_x,k_y)\right|^2}{\omega^2(k)-V^2k_x^2+2i\varepsilon Vk_x}\ . \label{WR}
\end{eqnarray}
Changing  to dimensionless variables consistently with the previous section, switching to polar coordinates, and inserting the dimensionless elliptical Gaussian pressure field defined in Eq.~(\ref{Gauss}) into Eq.~(\ref{WR}) leads to:
\begin{eqnarray}
R&=&  \lim_{\tilde \varepsilon\rightarrow 0}  \left( \frac{f_0^2}{4\pi^2 \rho gL^3}     \int \!\!\!\! \int \mathrm d K \mathrm d \theta\, \frac{iK^2 \cos\theta \exp\left(-\frac{K^2}{2\pi^2} \left(\cos^2\theta + \mathcal W^2 \sin^2\theta \right) \right)}{1-Fr^2K\cos^2\theta+2i\tilde \varepsilon Fr\cos\theta}\right) \ . \label{WR2}
\end{eqnarray}
Performing the integral over $K$ in Eq.~(\ref{WR2}) yields:
\begin{eqnarray}
R&=&\frac{f_0^2}{2\pi \rho gL^3}\,f(\mathcal V,\mathcal W) \ ,\label{res1}
\end{eqnarray}
where:
\begin{eqnarray}
f(\mathcal V, \mathcal W)&=&\frac{\mathcal W^3}{\mathcal V^6}\int_{0}^{\pi/2}\frac{\mathrm  d \theta}{\cos^5\theta}\,{\exp\left(-\frac{\mathcal W^2(1+\mathcal W^2 \tan^2\theta)}{2\pi^2\mathcal V^4\cos^2\theta}\right)}\ . \label{res2}
\end{eqnarray}
Equations (\ref{res1}) and (\ref{res2}) are central as they give the wave resistance of an elliptical Gaussian pressure field as a function of the dimensionless parameters $\mathcal V$ and $\mathcal W$. In order to seek for a limit form of the wave resistance at low aspect ratios, and guided by the fact that the integrand is dominant in the vicinity of $\pi/2$, this being all the more true for large values of $\mathcal V$ and small values of $\mathcal W$, we develop the integrand in the vicinity of $\pi/2$ at the lowest relevant order in $\theta$ and let the change of variables $\alpha=\pi/2-\theta$. This yields:
\begin{eqnarray}
f(\mathcal V, \mathcal W)&\simeq&\frac{\mathcal W^3}{\mathcal V^6}\int_{0}^{\pi/2}\frac{\mathrm  d \alpha}{\alpha^5}\,\exp\left(-\frac{\mathcal W^2}{2\pi^2\mathcal V^4 }\left(  \frac{\mathcal W^2}{\alpha^4}+\frac{1}{\alpha^2}\right)\right)\ .
\end{eqnarray}
Furthermore, aiming for a Gaussian like analytical integral, we let the change of variables $m=1/\alpha^2$. This leads to:
\begin{eqnarray}
f(\mathcal V, \mathcal W)&\simeq&\frac{\mathcal W^3}{2\mathcal V^6}\int_{4/\pi^2}^{\infty}{\mathrm  d m}\,m\,\exp\left(-\frac{\mathcal W^2}{2\pi^2\mathcal V^4 }\left( {\mathcal W^2}m^2+m\right)\right)\nonumber \\
&\simeq&\frac{  e^{ -\frac{2\mathcal W^2(\pi^2+4\mathcal W^2)}{\pi^6\mathcal V^4} }   }{8\mathcal V^4\mathcal W}\left[  4\pi^2\mathcal V^2-\pi^{3/2}\sqrt 2 \, e^{ \frac{(\pi^2+8\mathcal W^2)^2}{8\pi^6\mathcal V^4} }   \mathrm{erfc} \left(\frac{\pi^2+8\mathcal W^2}{2\pi^3\sqrt{2}\mathcal V^2}\right)\right] \ , \label{erfc}
\end{eqnarray}
where $\mathrm{erfc} =1-\mathrm{erf}$ is the complementary error function \citep{Abramowitz2012}. In the limit of small aspect ratios $\mathcal W$, $f(\mathcal V, \mathcal W)$ scales as $1/\mathcal W$ and:
\begin{eqnarray}
\lim_{\mathcal W\rightarrow 0}\mathcal W\,f(\mathcal V, \mathcal W)
&=& \frac{\pi^2}{2\mathcal V^2}-\frac{\pi^{3/2}\sqrt 2}{8\mathcal V^4} \, \exp\left({ \frac{1}{8\pi^2\mathcal V^4} }\right)   \mathrm{erfc} \left(\frac{1}{2\pi \sqrt{2}\mathcal V^2}\right)\ .\label{lim}
\end{eqnarray}
Equation (\ref{lim}) provides an analytical  limit of the wave resistance at very low aspect ratios $\mathcal W$. The fact that this limit is independent of $\mathcal W$ shows that $\mathcal V$ is the relevant variable to describe the problem of anisotropic moving objects. Figure \ref{resvag} displays $\mathcal W\,f(\mathcal V, \mathcal W)$ as given by Eq. (\ref{erfc})  as a function of  $\mathcal V$  for different values of the aspect ratio $\mathcal W$. The red dashed curve corresponds to the limit regime $\mathcal W \rightarrow 0$ as given analytically  by Eq.~(\ref{lim}). As one can see, the curves converge to the limit regime $\mathcal W\rightarrow 0$ as the aspect ratio is decreased, the convergence being faster at large $\mathcal V$. In the limit of small aspect ratios, the position of the maximum of wave resistance $\mathcal V_{\mathrm{max}}$ is obtained numerically by solving $\frac{\mathrm d}{\mathrm d\mathcal V}   \left( \lim_{\mathcal W\rightarrow 0}\mathcal W\,f(\mathcal V, \mathcal W)\right)=0$; one gets $\mathcal V_{\mathrm{max}}\simeq 0.3702$.  Recalling that $\mathcal V = Fr\sqrt{\mathcal W}$, one has the scaling of the position of the maximum of wave resistance in terms of Froude number: 
\begin{eqnarray}
Fr_{\mathrm{max}} \sim \frac{1}{\sqrt{\mathcal W}} \ ,
\end{eqnarray}
or identically in terms of real speed:
\begin{eqnarray}
V_{\mathrm{max}} \sim \sqrt{\frac{gL}{{\mathcal W}}}=\sqrt{gL}\,\sqrt{\frac{L}{{l}}} \ . \label{Vmax}
\end{eqnarray}
It is know that for cylindrical objects of size $L$, the position of the maximum of wave resistance scales as $\sqrt{gL}$ \citep{Stoker1992}. Equation (\ref{Vmax}) shows that in our case, the scaling is that of the cylindrical object multiplied by the square root of the inverse aspect ratio. Note that, in order to discuss the amplitude of wave resistance rather than the position of its maximum, one must prescribe how the total integrated force $f_0$ behaves with the dimensions of the distribution. For example, one could have $f_0$  proportional to the area $Ll=L^2\mathcal W$, and given that the wave resistance behaving as $f_0^2$, the relevant quantity to plot would be $\mathcal W^2\,f(\mathcal V, \mathcal W)$.

\section{Conclusion}\label{sec:conclusion}
In this study, we performed a theoretical analysis of the wake pattern and the wave resistance of an anisotropic moving disturbance. In order to account for anisotropy in a simple manner, we modelled the disturbance by an elliptical pressure distribution. In section 2, we derived the expression of the surface displacement and analysed it as a function of two relevant dimensionless parameters. We showed that  the angle delimiting the wake region remains constant and equal to the Kelvin angle $\varphi_\mathrm K = 19.47^{\circ}$; and that, at large Froude numbers, the angle corresponding to the maximum amplitude of the waves scales as $\sqrt{\mathcal W}/Fr$ where $\mathcal W$ is the aspect ratio of the elliptical disturbance and $Fr$ is the longitudinal hull Froude number. This notably extends the results of our previous study \citep{Darmon2014} that focused on isotropic moving objects. In section 3, we derived the expression of the wave resistance for an elliptical Gaussian pressure field and analysed it as a function of the same relevant dimensionless parameters. We obtained a limit analytical expression for the case of very small aspect ratios, and showed that the position of the maximum of wave resistance in terms of real speed scales as that of a cylindrical object multiplied by the square root of the inverse aspect ratio.
We believe this study is of interest as it reveals the main physical features of waves created by anisotropic moving objects, closer to the real geometry of ships than axisymmetric objects.

%

\section{Acknowledgements}\label{sec:acknowledgements}
We wish to thank A. C. Maggs for fruitful discussions.

\bibliographystyle{jfm}

\bibliography{biblio}

\end{document}